\documentclass[prb,onecolumn,preprint,showpacs,amsfonts,amssymb]{revtex4}%
\usepackage{amsfonts}
\usepackage{rotating}
\usepackage{amsmath}
\usepackage{amssymb}
\usepackage{graphicx}%
\providecommand{\U}[1]{\protect\rule{.1in}{.1in}}
\begin{document}
\preprint{ }
\title{Spin waves in magnetic quantum wells with Coulomb interaction and $sd$
exchange coupling}
\author{F. Perez}
\affiliation{Institut des NanoSciences de Paris, CNRS/Universit\'{e} Paris 6, 140 rue de
Lourmel, 75015 Paris, France}
\author{J. Cibert}
\affiliation{Institut N\'{e}el, CNRS/UJF, BP166, 38042 Grenoble cedex 9, France}
\author{M. Vladimirova and D. Scalbert}
\affiliation{Groupe d'Etude des Semiconducteurs, UMR 5650 CNRS, Universit\'{e} Montpellier
2, Place Eug\`{e}ne Bataillon, 34095 Montpellier cedex, France}

\begin{abstract}
We theoretically describe the spin excitation spectrum of a two dimensional
electron gas embedded in a quantum well with localized magnetic impurities.
Compared to the previous work, we introduce equations that
allow to consider the interplay between the Coulomb interaction of delocalized
electrons and the $sd$ exchange coupling between electrons and magnetic
impurities. Strong qualitative changes are found : mixed waves propagate below
the single particle continuum, an anticrossing gap is open at a specific
wavevector and the kinetic damping due to the electron motion strongly influences the coupling strength between electrons
and impurities spins.

\end{abstract}

\maketitle

\section{\label{intro}Introduction}

Collective spin dynamics in dilute magnetic semiconductors (DMS) has recently
drawn lots of attention.\cite{BaratePRB10,UltrafastGaMnAs09, GilbertFerro,
Vladimirova08,PerakisPRL08} This field provides an insight into the origins of
carrier-induced ferromagnetism in semiconductors \cite{FerroDMSDietl,
KoenigPRLEPR03, MacDonaldPRLFerroDMS} and particular features in the spin
excitation spectrum \cite{Mauger83,MacDonaldPRLFerroDMS} due to the presence
of two spin sub-systems that are dynamically coupled by Coulomb-exchange
interaction: that of the itinerant carrier and that of the localized magnetic
impurities. The transverse spin excitation spectrum has been theoretically
found to be composed of three types of excitations. These are: two collective
spin waves corresponding to itinerant and localized spins precessing in phase
or out of phase to each other, and single-particle (or Stoner-like)
excitations of the itinerant
carriers.\cite{Mauger83,MacDonaldPRLFerroDMS,Frustiglia04,Vladimirova08}. If
the DMS is in the ferromagnetic state, the in-phase spin wave (IPW) becomes
the Goldstone-like mode with an acoustic type dispersion responsible for
long-range spin order in the ground state. The out of phase spin wave (OPW)
develops an optical branch, with a zone-center energy determined by the
strength of Coulomb-exchange interaction between carriers and the spins of
magnetic impurities.

Experimental evidence of the entire spectrum in a ferromagnetic DMS like
GaMnAs is not available. What has been reported so far are features related to
the zone-center IPW, dominated by the Mn spin precession, its
dynamics.\cite{DynamicsGaMnAs,MagnonGaMnAs,UltrafastGaMnAs09} and its
ferromagnetic resonance.\cite{GaMnAsFMR} We find no experimental data
available for the out of phase mode. Indeed, ferromagnetism in GaMnAs systems
requires a high Mn concentration, which destroys the periodicity of the
crystal potential and smooths out all optical resonances.

More insight into the DMS spin excitation spectrum has been gained in CdMnTe
doped quantum wells (QW), which constitute a clean test-bed system,
appropriate to capture general properties of the collective spin dynamics in
DMS materials. Evidence for carrier-induced ferromagnetism has been found in
CdMnTe quantum wells doped with holes\cite{BoukariPRL02}. When doped with
electrons, due to the very low Curie temperature, only the paramagnetic phase
is available to most experiments. The OPW mode dispersion and single-particle
excitations have been probed by Raman measurements in the paramagnetic state
\cite{xCxsd,JusserandPRL03}. The mixed nature of the IPW and OPW waves has
been evidenced in the frequency\cite{Teran} and time
domain\cite{Vladimirova08,BaratePRB10}. Neithertheless, there is a lack of a
full theoretical description of the spin excitation spectrum in CdMnTe QW.
Indeed, so far two approaches were followed to describe the spin excitations :
in the first one\cite{Frustiglia04}, the $sd$-exchange dynamical coupling
between Mn and electron spins was considered, but the Coulomb interaction
between electrons was dropped out. In the second\cite{SpinReponse}, the
reverse point of view was adopted : spin resolved Coulomb interaction between
electrons was taken into account, but only the static mean-field $\mathit{sd}$
contribution from the Mn was kept to form a highly spin-polarized two
dimensional electron gas (SP2DEG).

This work fills the gap between the two theoretical approaches, by solving the
spin dynamical equations in presence of both the $sd$-exchange dynamical
coupling between Mn and electron spins and the Coulomb interaction between
electrons. Starting from the full DMS Hamiltonian, the approach combines exact
commutation rules and standard generalized Random Phase Approximation (RPA).
We also include the intrinsic damping of the pure electron spin waves due to
the delocalized character of the electrons\cite{Gomez,HankInhomo}. We show
that the introduction of Coulomb\ electron-electron interaction induces strong
qualitative changes in the spectrum compared to the approach of
Ref.\cite{Frustiglia04} and that inclusion of the intrinsic damping diminishes
the strength of the coupling between the two spin subsystems for non-zero
wavevectors. Generalization of this model to hole systems might be considered
: then one should take into account the fact that the hole spin states are not isotropic.

The paper is divided as follows : in Sec. II, we detail the Hamiltonian of the
system and rewrite it in terms of collective variables, in Sec. III, we use
transverse spin dynamics equations to derive spin response functions, and in
the last section, we study the spectrum of spin mixed electron-Mn modes.

\section{The 2DEG DMS Hamiltonian under static field}

We consider a QW of width $w$ containing $x_{eff}N_{0}$ unpaired\cite{Gaj} Mn
spins per unit volume. The first subband is populated by $n_{\text{2D}}$
electrons per unit surface. The 2DEG-DMS Hamiltonian under the influence of a
static magnetic field $\mathbf{B}=B\mathbf{e}_{z}$ applied in the plane of the
QW writes :

\bigskip%
\begin{gather}
\hat{H}=\hat{H}_{Kin}+\hat{H}_{Coulomb}+\hat{H}_{s-d}+\hat{H}_{Zeeman}%
\label{SP2DEGZeeman}\\
\hat{H}_{s-d}=-\alpha\iiint\mathbf{\hat{S}}\left(  \mathbf{r}\right)
\cdot\mathbf{\hat{M}}\left(  \mathbf{r}\right)  d^{3}r\nonumber\\
\hat{H}_{Zeeman}=g_{\text{e}}%
%TCIMACRO{\U{b5}}%
%BeginExpansion
\mu
%EndExpansion
_{\text{B}}\iiint\mathbf{\hat{S}}\left(  \mathbf{r}\right)  \cdot
\mathbf{B}d^{3}r+g_{\text{Mn}}%
%TCIMACRO{\U{b5}}%
%BeginExpansion
\mu
%EndExpansion
_{\text{B}}\iiint\mathbf{\hat{M}}\left(  \mathbf{r}\right)  \cdot
\mathbf{B}d^{3}r\nonumber
\end{gather}

\bigskip

where $\alpha$ is the exchange coupling between conduction electrons and Mn
spins $\left(  \alpha>0\right)  ,$ and $g_{e}$ and $g_{\text{Mn}}$ are normal
g-factor of, respectively, conduction electrons and Mn electrons. In the
convention where $%
%TCIMACRO{\U{b5}}%
%BeginExpansion
\mu
%EndExpansion
_{\text{B}}>0,$ we have $g_{\text{e}}\simeq-1.44$ and $g_{\text{Mn}}%
\simeq2.00$. We have introduced two vector operators : $\mathbf{\hat{S}%
}\left(  \mathbf{r}\right)  =\chi^{2}\left(  y\right)  \sum_{i}\mathbf{\hat
{s}}_{i}\delta\left(  \mathbf{r}_{//}-\mathbf{r}_{i//}\right)  $ is the 3D
electron spin density in a splitted coordinates frame $\mathbf{r=}\left(
\mathbf{r}_{//},y\right)  $ with $\mathbf{r}_{//},$ the in-plane position and
$y$ the out of plane coordinate. $\chi\left(  y\right)  $ is the electron
envelope-function of the first subband of the QW. The $i$ index accounts for
the $i$-th electron of the 2DEG, its spin $\frac{1}{2}$ is described by the
operator $\mathbf{\hat{s}}_{i}$ and its position is $\mathbf{r}_{i//}$.
$\mathbf{\hat{M}}\left(  \mathbf{r}\right)  =\sum_{j}\mathbf{\hat{I}}%
_{j}\delta\left(  \mathbf{r}-\mathbf{R}_{j}\right)  $ is the Mn 3D spin
density. The $j$-th $\frac{5}{2}$-spin $\mathbf{\hat{I}}_{j}$ of a single Mn
impurities is localized on the cation site $\mathbf{R}_{j}.$ In the
equilibrium state at temperature $T$, each Mn spin has the average value
$\left\langle \hat{I}_{z}\right\rangle \left(  B,T\right)  ,$ which is the
thermodynamic average over the five occupied states of the Mn atom d-shell,
given by the modified Brillouin function\cite{Gaj}. The 2DEG has the
equilibrium spin polarization $\zeta=\left(  n_{\uparrow}-n_{\downarrow
}\right)  /n_{\text{2D}}.$

We, now, rewrite the $s-d$ Hamiltonian using the electron (and Mn) spin
fluctuations operators at in-plane wave-vector $\mathbf{q}$, respectively,
$\mathbf{\hat{S}}_{\mathbf{q}}=\iiint\mathbf{\hat{S}}\left(  \mathbf{r}%
\right)  e^{-i\mathbf{q}\cdot\mathbf{r}_{//}}d^{2}r_{//}dy=\sum_{i}%
\mathbf{\hat{s}}_{i}e^{-i\mathbf{q}\cdot\mathbf{r}_{i//}}$ and $\mathbf{\hat
{M}}_{\mathbf{q}}=\sum_{j}\mathbf{\hat{I}}_{j}e^{-i\mathbf{q}\cdot
\mathbf{R}_{j//}}.$ Due to the 2D and 3D characters of, respectively, the
conduction electron and Mn spins subsystems, the electron spin-degrees of
freedom naturally couples to Mn spin profile weighted by the squared electron
wave-function. For later convenience, we introduce the following $n$-profile
Mn spin fluctuations operators :%

\[
\mathbf{\hat{M}}_{\mathbf{q}}^{(n)}=w^{n}\iiint\chi^{2n}\left(  y\right)
\mathbf{\hat{M}}\left(  \mathbf{r}\right)  e^{-i\mathbf{q}\cdot\mathbf{r}%
_{//}}d^{2}r_{//}dy=w^{n}\sum_{j}\chi^{2n}\left(  y_{j}\right)  \mathbf{\hat
{I}}_{j}e^{-i\mathbf{q}\cdot\mathbf{R}_{j//}}%
\]

Hence, $\mathbf{\hat{M}}_{\mathbf{q}}=$ $\mathbf{\hat{M}}_{\mathbf{q}}^{(0)}$.
$\mathbf{\hat{M}}_{\mathbf{q}}^{(n)}$ are vector operators verifying the
following commuting relations :%

\begin{equation}
\left[  \hat{M}_{\alpha,\mathbf{q}}^{(n)},\hat{M}_{\beta,\mathbf{q}^{\prime}%
}^{(p)}\right]  =i\epsilon_{\alpha,\beta,\gamma}\hat{M}_{\gamma,\mathbf{q}%
+\mathbf{q}^{\prime}}^{(n+p)} \label{nMnCommut}%
\end{equation}

where $\alpha,\beta,\gamma=x,y,z$ and $\epsilon_{\alpha,\beta,\gamma}$ is the
Levi-Cevita tensor. It follows :%

\begin{subequations}
\begin{align}
\hat{H}_{s-d}  &  =-\frac{\alpha}{wL^{2}}\sum_{\mathbf{q}}\mathbf{\hat{S}%
}_{\mathbf{q}}\cdot\mathbf{\hat{M}}_{-\mathbf{q}}^{(1)}\label{Hsdfull}\\
&  =-\tilde{\alpha}\left\{
\begin{array}
[c]{c}%
\delta\hat{S}_{z,\mathbf{q=0}}\cdot\left\langle \hat{M}_{z,\mathbf{q=0}}%
^{(1)}\right\rangle _{0}+\left\langle \hat{S}_{z,\mathbf{q=0}}\right\rangle
_{0}\cdot\delta\hat{M}_{z,\mathbf{q=0}}^{(1)}+\delta\hat{S}_{z,\mathbf{q=0}%
}\cdot\delta\hat{M}_{z,\mathbf{q=0}}^{(1)}\\
+\frac{1}{2}\hat{S}_{+,\mathbf{q=0}}\cdot\hat{M}_{-,\mathbf{q=0}}^{(1)}%
+\frac{1}{2}\hat{S}_{-,\mathbf{q=0}}\cdot\hat{M}_{+,\mathbf{q=0}}^{(1)}\\
+\sum_{\mathbf{q\neq0}}\mathbf{\hat{S}}_{\mathbf{q}}\cdot\mathbf{\hat{M}%
}_{-\mathbf{q}}^{(1)}%
\end{array}
\right\}  \label{Hsdfirst}%
\end{align}

In Eq. (\ref{Hsdfull}), we have defined the exchange coupling constant
$\tilde{\alpha}=\alpha/wL^{2}$, equilibrium averaging $\left\langle
{}\right\rangle _{0}$ and fluctuation operators : $\delta\hat{A}=\hat
{A}-\left\langle \hat{A}\right\rangle _{0}.$

Finally, we get :%

\end{subequations}
\begin{subequations}
\begin{align}
\hat{H}_{s-d}+H_{Zeeman}  &  =Z_{e}\left(  B\right)  \cdot\delta\hat
{S}_{z,\mathbf{q=0}}+g_{\text{Mn}}%
%TCIMACRO{\U{b5}}%
%BeginExpansion
\mu
%EndExpansion
_{B}B\cdot\delta\hat{M}_{z,\mathbf{q=0}}^{(0)}+K\cdot\delta\hat{M}%
_{z,\mathbf{q=0}}^{(1)}\label{HMF}\\
&  -\tilde{\alpha}\frac{1}{2}\left(  \hat{S}_{+,\mathbf{q=0}}\cdot\hat
{M}_{-,\mathbf{q=0}}^{(1)}-\hat{S}_{-,\mathbf{q=0}}\cdot\hat{M}%
_{+,\mathbf{q=0}}^{(1)}\right) \label{HmixedL}\\
&  -\tilde{\alpha}\delta\hat{S}_{z,\mathbf{q=0}}\cdot\delta\hat{M}%
_{z,\mathbf{q=0}}^{(1)}\label{Hdamp}\\
&  -\tilde{\alpha}\sum_{\mathbf{q\neq0}}\mathbf{\hat{S}}_{\mathbf{q}}%
\cdot\mathbf{\hat{M}}_{-\mathbf{q}}^{(1)} \label{Hcorrel}%
\end{align}

\bigskip

with the total bare Zeeman energy of conduction electrons given by :%

\end{subequations}
\begin{equation}
Z_{e}\left(  B\right)  =\underset{\Delta}{\underbrace{\tilde{\alpha}\gamma
_{1}N_{\text{Mn}}\left\vert \left\langle \hat{I}_{z}\right\rangle \left(
B,T\right)  \right\vert }}-\left\vert g_{\text{e}}\right\vert
%TCIMACRO{\U{b5}}%
%BeginExpansion
\mu
%EndExpansion
_{\text{B}}B \label{ZB0}%
\end{equation}

\bigskip

where $\gamma_{1}=\int_{0}^{w}\chi^{2}\left(  y\right)  dy$ is the probability
to find the electron in the QW, $N_{\text{Mn}}=x_{eff}N_{0}wL^{2}$ is the
number of Mn spins available in the QW. In Eq. (\ref{ZB0}), we evidence the
"Overhauser shift" $\Delta$ and the normal Zeeman contribution, opposite to
$\Delta$ (the $\mathit{sd}$ contribution). Indeed, as $g_{\text{e}%
}<0,g_{\text{Mn}}>0$ and $\alpha>0$, the Mn spins are anti-parallel to the
field, thus, the $sd$ coupling aligns the electron spins anti-parallel to the
field, while the normal Zeeman aligns the electron spin parallel to the field.
When the $sd$ coupling dominates over the normal Zeeman contribution and
$B>0$, both Mn and electron spins are $\downarrow$ :$\left\langle \hat{I}%
_{z}\right\rangle <0$ and the 2DEG spin polarization degree $\zeta$ is also negative.

The first line, Eq. (\ref{HMF}) gives the mean-field "effective Zeeman"
Hamiltonian, where no-dynamical coupling between electrons and Mn spins
appears. The mean-field Hamiltonian naturally introduces the "Knight shift",
due to the equilibrium electron spin polarization $\left\langle
S_{z,\mathbf{q=0}}\right\rangle _{0}=$ $n_{\text{2D}}L^{2}\zeta/2$, which
shifts the Mn spin precession energy:%

\begin{equation}
K=-\tilde{\alpha}\left\langle S_{z,\mathbf{q=0}}\right\rangle _{0}=\frac{1}%
{2}\frac{\alpha}{w}n_{\text{2D}}\left\vert \zeta\right\vert \label{Knight}%
\end{equation}

Eq. (\ref{HmixedL}) and Eq. (\ref{Hcorrel}) give rise to a first order
$sd-$dynamical coupling between transverse spin degrees of freedom and induce
spin-mixed electron-Mn modes of precession. They also contain higher orders
correlation terms which have been indentified to be responsible for a
damping\cite{PerakisPRL08}. The effects of the correlations contained in Eqs.
(\ref{HmixedL})-(\ref{Hcorrel}) are out of the scope of the present work and
will be neglected when they appear.

\section{Transverse spin dynamics}

\subsection{SP2DEG dynamics without the s-d dynamical coupling}

In this paragraph, we take into account only the first line of Eqs.(\ref{HMF}%
)-(\ref{Hcorrel}). Hence conduction electron and Mn dynamics are independent.
The 2DEG is polarized by the static exchange field of Mn spins, this forms a
spin-polarized 2DEG (SP2DEG) as described in Ref.\cite{SpinReponse}. The
Hamiltonian which rules the electron dynamics in the SP2DEG reduces to :%

\begin{equation}
\hat{H}_{SP2DEG}=\hat{H}_{Kin}+\hat{H}_{Coulomb}+Z_{e}\left(  B\right)
\hat{S}_{z,\mathbf{q=0}} \label{HSP2DEG}%
\end{equation}
Introducing the the electron creation-anihilation operators, a spin-flip
single particle excitation (SF-SPE) is described by a single electron-hole
pair operator $c_{\mathbf{k-q},\uparrow}^{+}c_{\mathbf{k},\downarrow}$, and
electrons spin-wave operators, introduced above, are given by $\hat
{S}_{+,\mathbf{q}}=\hat{S}_{x,\mathbf{q}}+i\hat{S}_{y,\mathbf{q}}%
=\sum_{\mathbf{k}}c_{\mathbf{k-q},\uparrow}^{+}c_{\mathbf{k},\downarrow}$.
Let's notice that $\left[  \hat{S}_{+,\mathbf{q}},c_{\mathbf{k-q}^{\prime
},\uparrow}^{+}c_{\mathbf{k},\downarrow}\right]  =0$, such that collective and
single particle modes are not intrinsically coupled. In the following we will
use exact commutation rules to write equation of motions for these normal
modes of the SP2DEG. We will note $\left(  d\hat{A}/dt\right)  _{2DEG}=\left[
\hat{A},\hat{H}_{SP2DEG}\right]  /i\hbar$ the time derivative of $\hat{A}$
related to $\hat{H}_{SP2DEG}$ only. Further, we will make use of linear
response theory to derive quantities like :
\begin{equation}
\left\langle \left\langle \hat{A};\hat{B}\right\rangle \right\rangle _{\omega
}=-\frac{i}{\hbar}\lim_{\varepsilon\rightarrow0^{+}}\int_{0}^{+\infty
}\left\langle \left[  \hat{A}\left(  t\right)  ,\hat{B}\right]  \right\rangle
_{0}e^{-i\omega t-\varepsilon t}dt \label{ABdef}%
\end{equation}

which gives the linear response of an observable $\hat{A}$ to a perturbation
coupled linearly to $\hat{B}$ in the considered
Hamiltonian\cite{QuantumLiquid}.

$\left\langle \left\langle \hat{A};\hat{B}\right\rangle \right\rangle
_{\omega}$ has the following equations of motion :%

\begin{subequations}
\begin{align}
\left\langle \left\langle \hat{A};\hat{B}\right\rangle \right\rangle
_{\omega}  &  =-\frac{i}{\omega}\left\langle \left\langle \dot{A};\hat
{B}\right\rangle \right\rangle _{\omega}-\frac{1}{\hbar\omega}\left\langle
\left[  \hat{A},\hat{B}\right]  \right\rangle _{0}\label{ABMotion}\\
&  =\frac{i}{\omega}\left\langle \left\langle \hat{A};\dot{B}\right\rangle
\right\rangle _{\omega}-\frac{1}{\hbar\omega}\left\langle \left[  \hat{A}%
,\hat{B}\right]  \right\rangle _{0} \label{ABMotionright}%
\end{align}

where $\dot{A}=\left[  \hat{A},\hat{H}\right]  /i\hbar$ is the time derivative
of $\hat{A}$ related to the considered Hamiltonian. We will note $\left\langle
\left\langle \hat{A};\hat{B}\right\rangle \right\rangle _{\omega}^{2DEG}$ when
$\dot{A}$ is replaced by $\left(  d\hat{A}/dt\right)  _{2DEG}$.

\subsubsection{Single particle modes dynamics}

The kinetic Hamiltonian $\hat{H}_{Kin}=\sum_{\mathbf{k,\sigma}}E_{\mathbf{k}%
}c_{\mathbf{k},\sigma}^{+}c_{\mathbf{k},\sigma}$ and the mean-field Zeeman
Hamiltonian conserve the single-particle modes :%

\end{subequations}
\begin{align*}
&  \left[  c_{\mathbf{k-q},\uparrow}^{+}c_{\mathbf{k},\downarrow},\hat
{H}_{Kin}+Z_{e}\left(  B\right)  \hat{S}_{z,\mathbf{q=0}}\right]  =\\
&  \left(  E_{\mathbf{k}}-E_{\mathbf{k-q}}-Z_{e}\right)  c_{\mathbf{k}%
-\mathbf{q},\uparrow}^{+}c_{\mathbf{k},\downarrow}%
\end{align*}

But the Coulomb Hamiltonian $\hat{H}_{Coulomb}=\frac{1}{2}\sum
\limits_{\mathbf{q}\neq0,\mathbf{k}^{\prime},\sigma^{\prime},\mathbf{k}%
,\sigma}V_{\mathbf{q}}c_{\mathbf{k+q},\sigma}^{+}c_{\mathbf{k}^{\prime
}-\mathbf{q},\sigma^{\prime}}^{+}c_{\mathbf{k}^{\prime},\sigma^{\prime}%
}c_{\mathbf{k},\sigma}$ couples a single particule mode to multi-pair modes
having a spin $+1$ :%

\begin{align}
&  \left[  c_{\mathbf{k-q},\uparrow}^{+}c_{\mathbf{k},\downarrow},\hat
{H}_{Coulomb}\right]  =\label{SPECoul}\\
&  +\sum\limits_{\mathbf{q}^{\prime}\neq0,\mathbf{k}^{\prime},\sigma
}V_{\mathbf{q}^{\prime}}\left(  c_{\mathbf{k-q+q}^{\prime},\uparrow}%
^{+}c_{\mathbf{k}^{\prime},\sigma}c_{\mathbf{k}^{\prime}-\mathbf{q}^{\prime
},\sigma}^{+}c_{\mathbf{k},\downarrow}\right) \nonumber\\
&  -\sum\limits_{\mathbf{q}^{\prime}\neq0,\mathbf{k}^{\prime},\sigma
}V_{\mathbf{q}^{\prime}}\left(  c_{\mathbf{k-q},\uparrow}^{+}c_{\mathbf{k}%
^{\prime},\sigma}c_{\mathbf{k}^{\prime}-\mathbf{q}^{\prime},\sigma}%
^{+}c_{\mathbf{k-q}^{\prime},\downarrow}\right) \nonumber
\end{align}

where $V_{\mathbf{q}^{\prime}}$ is the space Fourier transform of the bare
Coulomb interaction. It follows from Eq. (\ref{SPECoul}) that $\hat
{H}_{Coulomb}$ does not conserve the SPE motion, but introduces an infinite
hierarchy where a single electron-hole pair of the Fermi sea (a SPE) couples
to multiple pairs having the same global spin. Approximations can be made : in
the rhs terms of Eq. (\ref{SPECoul}), some conserve the SPE motion and
renormalize it, others introduce a scattering effect, the so-called
spin-Coulomb drag\cite{DamicoEuro01}, and can be described by an
electron-electron scattering time$\cite{Hank1}$ $\tau_{e-e}.$ The former
consists in making the random phase approximation (RPA) on single mode
dynamics\cite{IzuyamaJPS63},\textit{ i.e.}, keeping in Eq. (\ref{SPECoul})
only terms which can be written as a product of a SF-SPE with an occupation
number $\hat{n}_{\mathbf{k},\sigma}$, and replacing $\hat{n}_{\mathbf{k}%
,\sigma}$ by its average value $\left\langle \hat{n}_{\mathbf{k},\sigma
}\right\rangle _{0}$. Then the Coulomb factor $V_{\mathbf{q}^{\prime}}$ has to
be replaced by a local field factor $G_{xc}$ which accounts for the effective
dynamical exchange-field produced by other electrons\cite{MarinescuPRB97} (a
part of what has been suppressed in making the RPA). Adding a damping rate
$\eta=\hbar/\tau_{e-e}$ due to the scattering leads to the SF-SPE equation of
motion that we will use in the following :%

\begin{gather}
i\hbar\left(  \frac{d}{dt}c_{\mathbf{k-q},\uparrow}^{+}c_{\mathbf{k}%
,\downarrow}\right)  _{2DEG}=\left(  E_{\mathbf{k}}-E_{\mathbf{k-q}}%
-Z_{e}+2G_{xc}\left\langle \hat{S}_{z,\mathbf{q=0}}\right\rangle _{0}\right)
c_{\mathbf{k}-\mathbf{q},\uparrow}^{+}c_{\mathbf{k},\downarrow}%
\label{SFSPEdyna}\\
+G_{xc}\left(  \left\langle \hat{n}_{\mathbf{k},\downarrow}\right\rangle
_{0}-\left\langle \hat{n}_{\mathbf{k}-\mathbf{q},\uparrow}\right\rangle
_{0}\right)  \hat{S}_{+,\mathbf{q}}-i\eta c_{\mathbf{k-q},\uparrow}%
^{+}c_{\mathbf{k},\downarrow}\nonumber
\end{gather}
Eq. (\ref{SFSPEdyna}) evidences the renormalized Zeeman energy, \textit{i.e.},
the spin-flip energy of single electrons :%

\begin{equation}
Z^{\ast}=Z_{e}-2G_{xc}\left\langle \hat{S}_{z,\mathbf{q=0}}\right\rangle _{0}
\label{Zstar}%
\end{equation}

Compared to the bare Zeeman energy $Z_{e}$, $Z^{\ast}$ is enhanced by
Coulomb-exchange between spin-polarized electrons, a phenomenon linked to the
spin-susceptibility enhancement\cite{Spinsusc}. Each SF-SPE is characterized
by two wavevectors $\mathbf{k}$ and $\mathbf{q}$. At $\mathbf{q}=0$, SF-SPE
are degenerate to $Z^{\ast}.$ When $\mathbf{q}\neq0$ the degeneracy is lifted
by the kinetic spread of velocities which depend on the initial momentum
$\mathbf{k}$.

\subsubsection{Collective modes dynamics}

Along the SF-SPE, the above spin polarized SP2DEG develops collective modes,
the so-called spin-flip waves (SFW). SFW dynamics are described by the
$\hat{S}_{+,\mathbf{q}}$ operators dynamics. As $\hat{H}_{Coulomb}$ conserves
the macroscopic spin, it follows :%

\begin{equation}
\left[  \hat{S}_{+,\mathbf{q}},\hat{H}_{Coulomb}+Z_{e}\left(  B\right)
\hat{S}_{z,\mathbf{q=0}}\right]  =-Z_{e}\left(  B\right)  \hat{S}%
_{+,\mathbf{q}} \label{S+QZH}%
\end{equation}

But, the kinetic Hamiltonian couples collective states to the spin current
$\mathbf{\hat{J}}_{+,\mathbf{q}}=\frac{\hbar}{m^{\ast}}\sum_{\mathbf{k}%
}\left(  \mathbf{k-}\frac{\mathbf{q}}{2}\right)  c_{\mathbf{k-q},\uparrow}%
^{+}c_{\mathbf{k},\downarrow}$ carried by single particle states :%

\begin{equation}
\left[  \hat{S}_{+,\mathbf{q}},\hat{H}_{Kin}\right]  =\hbar\mathbf{q\cdot
\hat{J}}_{+,\mathbf{q}} \label{S+QHKin}%
\end{equation}

Finally the equation of motion of collective SFW modes writes :%

\begin{equation}
\left(  \frac{d}{dt}\hat{S}_{+,\mathbf{q}}\right)  _{2DEG}=i\omega_{e}\hat
{S}_{+,\mathbf{q}}-i\mathbf{q\cdot\hat{J}}_{+,\mathbf{q}} \label{S+QMotion}%
\end{equation}
where $\omega_{e}=Z_{e}\left(  B\right)  /\hbar$ is the frequency of the
Larmor's electron mode.

One is left with evaluating the spin-current dynamics to find the 2DEG
electron spin waves. The spin current evolution is dominated by single
particle states dynamics as $\hat{H}_{Coulomb}$ does not conserve
$\mathbf{\hat{J}}_{+,\mathbf{q}}$ and destroys the coherence between the
single particle objects composing $\mathbf{\hat{J}}_{+,\mathbf{q}}.$ As seen
from Eq. (\ref{SFSPEdyna}), the exchange field produced by the spin
fluctuation $\hat{S}_{+,\mathbf{q}}$ drives the spin current. The interplay
between the spin-current dynamics and the $\hat{S}_{+,\mathbf{q}}$ dynamics
then determines both the SFW dispersion and its damping. The relevant
spin-current response is the transverse spin-conductivity\cite{HankInhomo}
$\tilde{\sigma}_{+},$ which links the spin current to the gradient of the
exciting exchange field :%

\begin{equation}
\left\langle \mathbf{\hat{J}}_{+,\mathbf{q}}\right\rangle _{\omega}%
=\mathbf{q}\tilde{\sigma}_{+}\left(  \mathbf{q},\omega\right)  G_{xc}%
\left\langle \hat{S}_{+,\mathbf{q}}\right\rangle _{\omega} \label{J+qS+q}%
\end{equation}
where $\left\langle {}\right\rangle _{\omega}$ is the expectation value at
frequency $\omega.$ The spin-conductivity has an imaginary part originating
from the damping of SF-SPE, intrinsically due to $\hat{H}_{Coulomb}$ or any
source of disorder acting on transverse spin degrees of freedom. Consequently,
the real part of the spin conductivity determines the SFW dispersion, while
the imaginary part determines its damping. It is worth noting, that the spin
wave damping originates from the kinetic motion of the conduction electrons
and from the topoly of the conduction band, a 2D parabolla. In a Luttinger
liquid\cite{LuttingerReview}, the conduction band is linear in $\mathbf{k}$
and 1D, thus Eq. (\ref{S+QHKin}) conserves the macroscopic spin. It breaks the
coupling between $\hat{S}_{+,\mathbf{q}}$ and SF-SPEs, which are coupled to
charge degrees of freedom by $\hat{H}_{Coulomb}$. This property is at the
origin of the well known spin-charge separation\cite{Spincharge} occuring in
Luttinger liquids. Injecting Eq. (\ref{J+qS+q}) into Eq. (\ref{S+QMotion}) and
solving it in the frequency domain for long wavelength ($q\ll k_{F}$) leads to :%

\begin{equation}
\frac{d}{dt}\hat{S}_{+,\mathbf{q}}=i\tilde{\omega}_{q}\hat{S}_{+,\mathbf{q}}
\label{S+qapprox}%
\end{equation}

with $\tilde{\omega}_{q}$ a complex pulsation:%

\begin{subequations}
\begin{align}
\operatorname{Re}\tilde{\omega}_{q}  &  =\omega_{e}-q^{2}G_{xc}\lim
_{q\rightarrow0,\omega\rightarrow0}\operatorname{Re}\tilde{\sigma}%
_{+}\label{omegaqRe}\\
\operatorname{Im}\tilde{\omega}_{q}  &  =q^{2}G_{xc}\lim_{q\rightarrow
0,\omega\rightarrow0}\operatorname{Im}\tilde{\sigma}_{+} \label{omegaqIm}%
\end{align}

In the following we note $\sigma_{+}=\lim_{q\rightarrow0,\omega\rightarrow
0}\operatorname{Im}\tilde{\sigma}_{+}$, the imaginary spin-conductivity. It
was calculated in Ref.\cite{HankInhomo} and some corrections were added in
Ref.\cite{Gomez} which gave also an experimental evidence of the kinetic
damping law found in Eq. (\ref{omegaqIm}). We highlight that these $q^{2}$
laws are valid in the longwavelength limit when the SFW propagates far from
the SF-SPE continuum (see Ref.\cite{SpinReponse}). When close to this
continuum, the stronger coupling with SF-SPEs introduces corrections to the
above laws and one should better replace the dispersion law with the pole
appearing in the transverse spin susceptibility (see Ref.\cite{SpinReponse})
which will be derived in the next paragraph.

\subsubsection{Electron spin-susceptibility}

The transverse spin-susceptibility, defined by the ratio of the expectation
value $\left\langle \hat{S}_{+,\mathbf{q}}\right\rangle _{\omega}$ to the
perturbing potentiel $g_{e}%
%TCIMACRO{\U{b5}}%
%BeginExpansion
\mu
%EndExpansion
_{B}b_{+,\mathbf{q\omega}}$, where $b_{+,\mathbf{q\omega}}$ is the amplitude
at the same pulsation $\omega$ and wavevector $\mathbf{q}$ of the exciting
magnetic field, is given by :%

\end{subequations}
\begin{equation}
\chi_{+}\left(  \mathbf{q,\omega}\right)  =\left\langle \left\langle \hat
{S}_{+,\mathbf{q}};\hat{S}_{-,-\mathbf{q}}\right\rangle \right\rangle
_{\omega}^{2DEG} \label{Spinsuscdef}%
\end{equation}

Straightforward calculations using the equation of motion Eq. (\ref{SFSPEdyna}%
) lead to :%

\begin{equation}
\chi_{+}\left(  \mathbf{q,\omega}\right)  =\left\langle \left\langle \hat
{S}_{+,\mathbf{q}};\hat{S}_{-,-\mathbf{q}}\right\rangle \right\rangle
_{\omega}^{2DEG}=\frac{\Pi_{\downarrow\uparrow}\left(  \mathbf{q}%
,\omega\right)  }{1+G_{xc}\Pi_{\downarrow\uparrow}\left(  \mathbf{q}%
,\omega\right)  } \label{Spinsusc}%
\end{equation}

where we have introduced the transverse Lindhardt-type
response\cite{SpinReponse} :%

\begin{equation}
\Pi_{\downarrow\uparrow}\left(  \mathbf{q},\omega\right)  =\sum_{\mathbf{k}%
}\frac{\left\langle \hat{n}_{\mathbf{k-q},\uparrow}\right\rangle
_{0}-\left\langle \hat{n}_{\mathbf{k},\downarrow}\right\rangle _{0}}{Z^{\ast
}+E_{\mathbf{k-q}}-E_{\mathbf{k}}-\hbar\omega-\eta} \label{Piupdn}%
\end{equation}

A comparison between the above spin-susceptibility expression and the one
given by local spin-density approximation\cite{SpinReponse}, gives the
expression of the local field factor $G_{xc}:$%

\begin{equation}
G_{xc}=\mathbf{-}\frac{2}{n_{2D}^{2}L^{2}}\frac{1}{\zeta}\frac{\partial
E_{xc}}{\partial\zeta} \label{Gxc}%
\end{equation}

where $E_{xc}$ is the exchange-correlation part of the ground state
energy\cite{GoriGiorgi}.

SFW appear as poles of $\chi_{+}\left(  \mathbf{q,\omega}\right)  ,$ one finds
in the long wavelength limit, another expression for $\operatorname{Re}%
\tilde{\omega}_{q}$ :%

\begin{equation}
\operatorname{Re}\tilde{\omega}_{q}=\omega_{e}-\frac{1}{\left\vert
\zeta\right\vert }\frac{Z_{e}}{Z^{\ast}-Z_{e}}\frac{\hbar}{2m^{\ast}}q^{2}
\label{omegaq}%
\end{equation}

Alternatively, if one uses the approximated equation of motion Eq.
(\ref{S+QMotion}), one finds the spin susceptibility in the long wavelength
limit :%

\begin{equation}
\chi_{+}\left(  \mathbf{q,\omega}\right)  =-\frac{2\left\langle \hat
{S}_{z,\mathbf{q=0}}\right\rangle _{0}}{\hbar\omega-\hbar\tilde{\omega}_{q}}
\label{Spinsuscapprox}%
\end{equation}

\subsection{\label{SP2DEG}Transverse spin dynamics equations with s-d
dynamical coupling}

Now, we keep lines (\ref{HmixedL}) to (\ref{Hcorrel}) in the s-d Hamiltonian,
and we reconsider collective transverse spin dynamics. In the following, the
derivative $d\hat{A}/dt=\left[  \hat{A},\hat{H}\right]  /i\hbar$ takes into
account the coherent coupled dynamics due to lines (\ref{HmixedL}) and
(\ref{Hcorrel}) in the s-d Hamiltonian, but reduced to first order terms :
higher order correlation terms like $\sum_{\mathbf{q}^{\prime}}\delta\hat
{S}_{z,\mathbf{q+q}^{\prime}}\cdot\hat{M}_{+,-\mathbf{q}^{\prime}}^{(1)}$ have
been dropped$.$

\subsubsection{Electron dynamics\bigskip}

We find :%

\begin{equation}
\frac{d}{dt}\hat{S}_{+,\mathbf{q}}=\dot{S}_{+,\mathbf{q}}-\frac{i}{\hbar}%
K\hat{M}_{+,\mathbf{q}}^{(1)} \label{ElecDynaMixed}%
\end{equation}

where $\dot{S}_{+,\mathbf{q}}=i\tilde{\omega}_{q}\hat{S}_{+,\mathbf{q}}.$
Compared to the SP2DEG dynamics, the $sd$-dynamical coupling adds the second
term of Eq. (\ref{ElecDynaMixed}) which is a coherent coupling with Mn
transverse degrees of freedom. One key feature is that the collective electron
motion naturally couples with $\hat{M}_{+,\mathbf{q}}^{(1)}$ Mn-modes, a Mn
precession having a profile, out of the QW plane, following the electron
probability distribution. We are left with deriving the equation of motion for
these Mn-modes.

\subsubsection{Manganese dynamics}

We obtain the first order equation of motion for Mn spins :%

\begin{equation}
\frac{d}{dt}\hat{M}_{+,\mathbf{q}}^{(n)}=\frac{i}{\hbar}g_{Mn}%
%TCIMACRO{\U{b5}}%
%BeginExpansion
\mu
%EndExpansion
_{B}B\hat{M}_{+,\mathbf{q}}^{(n)}+\frac{i}{\hbar}K\hat{M}_{+,\mathbf{q}%
}^{(n+1)}-\frac{i}{\hbar}\Delta_{n+1}\hat{S}_{+,\mathbf{q}}
\label{Mndynamixed}%
\end{equation}

where we have introduced $n$-profile Overhauser shifts : $\Delta_{n}%
=\tilde{\alpha}\left\vert \left\langle \hat{M}_{z,\mathbf{q=0}}^{(n)}%
\right\rangle _{0}\right\vert =\gamma_{n}/\gamma_{1}\Delta$ with $\gamma
_{n}=w^{n-1}\int_{0}^{w}\chi^{2n}\left(  y\right)  dy.$ The important features
are the\ second and third terms in Eq. (\ref{Mndynamixed}). The later couples
the Mn-precession with collective electron modes. The former couples a
$n$-profile Mn mode to a $\left(  n+1\right)  $-profile mode, because this
coupling is mediated by the 2DEG. Thus, the Mn-dynamics is given by an
infinite serie of equations. This is a consequence of the 3D nature of the Mn
dynamics. A variable like $\hat{M}_{+,\mathbf{q}}^{(n)}$ describes an
oscillation propagating in the plane with a rigid profile in the normal
direction, but the out of plane degree of freedom is restored by the
possibility for Mn spins to build modes which are combinations of $\hat
{M}_{+,\mathbf{q}}^{(n)}$ resulting in different out of plane
profiles\cite{Frustiglia04}. Obviously, the $\hat{M}_{+,\mathbf{q}}^{(n)}$ are
not independent variables because they don't correspond to orthogonal out of
plane profiles. Solving the serie of infinite equations requires a projection
of $\hat{M}_{+,\mathbf{q}}^{(n)}$ over a set of modes with orthogonal profiles
as it was carried out in Ref.\cite{Frustiglia04}. Along with modes having a
strong mixed nature (electron-Mn modes), we then expect to find a high number
of modes having essentially a Mn character, but with orthogonal profiles (Mn
modes). The number of Mn modes has to be consistent with the initial number of
degrees of freedom present in the system. Ref. \cite{Frustiglia04} found a
high number of Mn modes branches which were separated by energies of the order
of 0.1$\Delta.$ However, in the experimental data of Ref.\cite{Vladimirova08},
only one branch of these Mn modes was apparent. It appears then, that the set
of modes chosen in Ref.\cite{Frustiglia04} is not the most appropriate to
describe properly all the modes contained in Eqs.(\ref{ElecDynaMixed}%
)-(\ref{Mndynamixed}), at least in the vicinity of the anticrossing gap (see
below). Anyway, this point requires further developments out of the scope of
the present study. Indeed, we are particularly interested in discussing mixed
electron-Mn modes which are strongly coupled to electrons rather than modes
specific to the 3D nature of the Mn dynamics. We can remark that the coupling
between $\hat{M}_{+,\mathbf{q}}^{(n)}$ and $\hat{M}_{+,\mathbf{q}}^{(n+1)}$
has a strength given by $K,$ which is very small compared to $\Delta_{n}$ due
to the ratio $\left\vert \left\langle \hat{M}_{z,\mathbf{q=0}}^{(n)}%
\right\rangle _{0}/\left\langle \hat{S}_{z,\mathbf{q=0}}\right\rangle
_{0}\right\vert \gg1.$ Hence, considering only modes strongly coupled with
electron modes is reasonable. As electron modes are naturally coupled to
$\hat{M}_{+,\mathbf{q}}^{(1)}$ modes, we will consider the dynamics for these
ones only by cutting the infinite serie with an homothetic approximation :%

\begin{equation}
\hat{M}_{+,\mathbf{q}}^{(2)}=\left(  \gamma_{2}/\gamma_{1}\right)  \hat
{M}_{+,\mathbf{q}}^{(1)} \label{homothetic}%
\end{equation}

Consequently the set of coupled electron-Mn equations reduces to :%

\begin{subequations}
\begin{align}
\frac{d}{dt}\hat{S}_{+,\mathbf{q}}  &  =i\tilde{\omega}_{q}\hat{S}%
_{+,\mathbf{q}}-\frac{i}{\hbar}K\hat{M}_{+,\mathbf{q}}^{(1)}%
\label{Elecdynafinal}\\
\frac{d}{dt}\hat{M}_{+,\mathbf{q}}^{(1)}  &  =i\omega_{Mn}\hat{M}%
_{+,\mathbf{q}}^{(1)}-\frac{i}{\hbar}\Delta_{2}\hat{S}_{+,\mathbf{q}}
\label{Mndynafinal}%
\end{align}

where :%

\end{subequations}
\begin{equation}
\omega_{Mn}=\left(  g_{Mn}%
%TCIMACRO{\U{b5}}%
%BeginExpansion
\mu
%EndExpansion
_{B}B+K\frac{\gamma_{2}}{\gamma_{1}}\right)  /\hbar\label{Mnpuls}%
\end{equation}

is the natural precession pulsation of the free $\hat{M}_{+,\mathbf{q}}^{(1)}$ mode.

\section{Mixed Mn-electron spin waves}

\subsection{Spin susceptibilities}

To find the dynamically coupled modes, we will derive the electron spin
susceptibility with help of equations of motion (\ref{ABMotion}) and
(\ref{Elecdynafinal})-(\ref{Mndynafinal}).

From Eqs.(\ref{ABMotion}) and (\ref{Elecdynafinal}), we first get :%
\[
\left\langle \left\langle \hat{S}_{+,\mathbf{q}};\hat{S}_{-,-\mathbf{q}%
}\right\rangle \right\rangle _{\omega}=\frac{\tilde{\omega}_{q}}{\omega
}\left\langle \left\langle S_{+,\mathbf{q}};\hat{S}_{-,-\mathbf{q}%
}\right\rangle \right\rangle _{\omega}-\frac{K}{\hbar\omega}\left\langle
\left\langle \hat{M}_{+,\mathbf{q}}^{(1)};\hat{S}_{-,-\mathbf{q}}\right\rangle
\right\rangle _{\omega}-\frac{2}{\hbar\omega}\left\langle \hat{S}%
_{z,\mathbf{q=0}}\right\rangle _{0}%
\]

then from Eq. (\ref{Mndynafinal}), we get :%

\[
\left\langle \left\langle \hat{M}_{+,\mathbf{q}}^{(1)};\hat{S}_{-,-\mathbf{q}%
}\right\rangle \right\rangle _{\omega}=\frac{\omega_{Mn}}{\omega}\left\langle
\left\langle \hat{M}_{+,\mathbf{q}}^{(1)};\hat{S}_{-,-\mathbf{q}}\right\rangle
\right\rangle _{\omega}-\frac{\Delta_{2}}{\hbar\omega}\left\langle
\left\langle \hat{S}_{+,\mathbf{q}};\hat{S}_{-,-\mathbf{q}}\right\rangle
\right\rangle _{\omega}%
\]

hence,%

\[
\left\langle \left\langle \hat{M}_{+,\mathbf{q}}^{(1)};\hat{S}_{-,-\mathbf{q}%
}\right\rangle \right\rangle _{\omega}=-\Delta_{2}\frac{\left\langle
\left\langle \hat{S}_{+,\mathbf{q}};\hat{S}_{-,-\mathbf{q}}\right\rangle
\right\rangle _{\omega}}{\hbar\omega-\hbar\omega_{Mn}}%
\]

which finally leads to :%

\begin{equation}
\left\langle \left\langle \hat{S}_{+,\mathbf{q}};\hat{S}_{-,-\mathbf{q}%
}\right\rangle \right\rangle _{\omega}=\frac{\left(  \hbar\omega-\hbar
\omega_{Mn}\right)  \chi_{+}\left(  \mathbf{q,\omega}\right)  }{\hbar
\omega-\hbar\omega_{Mn}-\frac{\tilde{\alpha}\Delta_{2}}{2}\chi_{+}\left(
\mathbf{q,\omega}\right)  } \label{Suscelec}%
\end{equation}

and%
\begin{equation}
\left\langle \left\langle \hat{M}_{+,\mathbf{q}}^{(1)};\hat{M}_{-,-\mathbf{q}%
}^{(1)}\right\rangle \right\rangle _{\omega}=\frac{-2\left\langle \hat
{M}_{z,\mathbf{q=0}}^{(1)}\right\rangle }{\hbar\omega-\hbar\omega_{Mn}%
-\frac{\tilde{\alpha}\Delta_{2}}{2}\chi_{+}\left(  \mathbf{q,\omega}\right)  }
\label{SuscMn}%
\end{equation}

Consequently, e-Mn mixed spin excitations appear as poles of the above
responses, \textit{i.e.}, are zeros of the propagator :%

\begin{equation}
\hbar\omega-\hbar\omega_{Mn}-\frac{\tilde{\alpha}\Delta_{2}}{2}\chi_{+}\left(
\mathbf{q,\omega}\right)  \label{Mnpoles}%
\end{equation}

with $\chi_{+}\left(  \mathbf{q,}\omega\right)  $ being the
spin-susceptibility of the uncoupled SP2DEG described in Section \ref{SP2DEG}.

We can understand the above equation as follows. Consider the Mn point of view
; in the presence of the SP2DEG, the precession frequency of Mn spins is
shifted from the normal precession ($g_{Mn}%
%TCIMACRO{\U{b5}}%
%BeginExpansion
\mu
%EndExpansion
_{B}B/\hbar$) by two quantities : a blue shift due to static exchange field
with spin polarized electrons ($K/\hbar)$ and an additional shift due to the
dynamic change of the electron spin-polarization. The later is induced by the
Mn precession itself. Finally Eq. (\ref{Mnpoles}) describes a recursive closed
loop where : Mn transverse precession induces electron transverse precession
proportional to $\Delta_{2}\chi_{+}\left(  \mathbf{q,}\omega\right)  ,$ this
dynamically changes the electron spin polarization which in turn shifts the Mn
precession frequency by an amount $\tilde{\alpha}\Delta_{2}\chi_{+}\left(
\mathbf{q,}\omega\right)  .$ Finally, in dropping correlation terms given by
Eqs. (\ref{Hdamp}) and (\ref{Hcorrel}), one finds a collective behavior where
electrons and Mn respond adiabatically to the dynamical perturbation from the
opposite spin-subsystem.

Similar expressions for the coupled modes propagator have been obtained in
previous works. To our knowledge it was first derived in Ref.\cite{Mauger83}
for bulk DMS, and more recently using a spin-path integral approach in DMS
quantum wells with electrons \cite{Frustiglia04,KoenigPRLEPR03} or bulk DMS
with holes\cite{MacDonaldPRLFerroDMS}. However, none of these works did
include the influence of the Coulomb interaction between carriers. Instead of
Eq. (\ref{Mnpoles}), they resulted in the following propagator :%

\begin{equation}
\hbar\omega-\hbar\omega_{Mn}-\frac{\tilde{\alpha}\Delta}{2}\Pi_{\downarrow
\uparrow}\left(  \mathbf{q,}\omega\right)  \label{NonIntProp}%
\end{equation}

where the electron spin-susceptibility was replaced by the non-interacting
single-particle response introduced in Eq. (\ref{Piupdn}). Introduction of
Coulomb interaction results in strong qualitative changes in the spectrum
which have been partially adressed in Ref.\cite{xCxsd} and will be detailed below.

\subsection{Homogeneous modes}%

%TCIMACRO{\FRAME{fhFU}{4.5723in}{3.218in}{0pt}{\Qcb{(Color online). Zone center
%electron-Mn modes. Dotted lines are the uncoupled electron (upper curve) and
%Mn (lower curve) modes. Full lines are the solutions $\omega_{q=0}^{\pm}$ of
%Eq.(\ref{ZoneCenterZeros}). Out of the resonant field $B_{\text{R}}$ where the
%modes anticross, branches have electron or Mn character. Sample parameters
%are, $x_{eff}=0.23\%,$ $T=2$ K, $w=150\mathring{A}$ and $n_{\text{2D}%
%}=3.1\times10^{11}$cm$^{-2}.$}}{\Qlb{FigAntiCross}}{figanticross.wmf}%
%maintain-aspect-ratio TRUE;  display "USEDEF";  valid_file "T";
%width 4.5723in;  height 3.218in;  depth 0pt;  original-width 4.5204in;
%original-height 3.1739in;  cropleft "0";  croptop "1";  cropright "1";
%cropbottom "0";  tempfilename 'FigAnticross.wmf';tempfile-properties "XNPR";}%
%}}%
%BeginExpansion
\begin{figure}
[h]
\begin{center}
\includegraphics[
natheight=3.173900in,
natwidth=4.520400in,
height=3.218in,
width=4.5723in
]%
{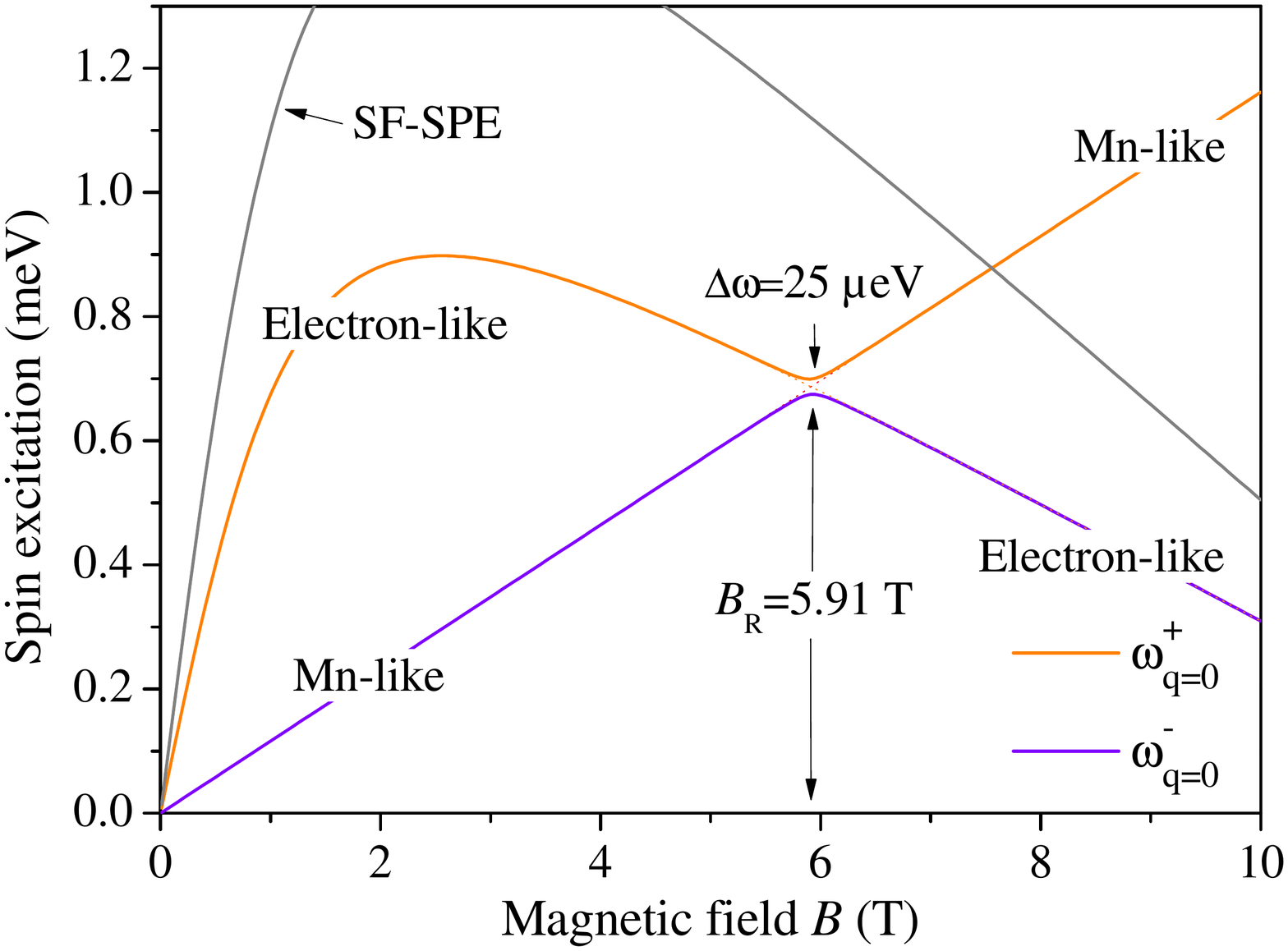}%
\caption{(Color online). Zone center electron-Mn modes. Dotted lines are the
uncoupled electron (upper curve) and Mn (lower curve) modes. Full lines are
the solutions $\omega_{q=0}^{\pm}$ of Eq.(\ref{ZoneCenterZeros}). Out of the
resonant field $B_{\text{R}}$ where the modes anticross, branches have
electron or Mn character. Sample parameters are, $x_{eff}=0.23\%,$ $T=2$ K,
$w=150\mathring{A}$ and $n_{\text{2D}}=3.1\times10^{11}$cm$^{-2}.$}%
\label{FigAntiCross}%
\end{center}
\end{figure}
%EndExpansion

Eq. (\ref{NonIntProp}) was used to successfully fit the experiment of Teran et
al. (Ref.\cite{Teran}) where homogenous modes ($q$=0) were probed and shown to
experience an anticrossing at a magnetic field $B_{R}$ such that $\omega
_{Mn}=\omega_{e}$. It is a consequence of the Larmor's theorem that the
homogeneous electron mode behaves as if electron were not interacting (Eq.
(\ref{S+QMotion}) for $q=0$). Indeed, setting $q=0$ in Eq. (\ref{Mnpoles}),
leads to the homogenous precession modes equation :%

\begin{equation}
\left(  \hbar\omega-\hbar\omega_{Mn}\right)  \left(  \hbar\omega-\hbar
\omega_{e}\right)  -K\Delta_{2}=0 \label{ZoneCenterZeros}%
\end{equation}

which solutions are real :%

\begin{equation}
\omega_{q=0}^{\pm}=\frac{\omega_{e}+\omega_{Mn}}{2}\pm\frac{1}{2}\sqrt{\left(
\omega_{e}-\omega_{Mn}\right)  ^{2}+4K\Delta_{2}/\hbar^{2}} \label{Homogpoles}%
\end{equation}

Figure~\ref{FigAntiCross} shows the magnetic field dependence of these modes
and the gap opening at the resonant field $B_{\text{R}}$. The upper branch
$\omega_{q=0}^{+}$ (resp $\omega_{q=0}^{-}$) has an electron character (resp.
Mn character) when $B<B_{\text{R}}$ and vice-versa for $B>B_{\text{R}}.$ The
amplitude of the homogeneous anticrossing gap $\Delta\omega_{q=0}%
=\sqrt{4K\Delta_{2}}$ denotes the strength of the dynamical coupling between
the two spin subsystems. Detailed discussions on this gap have been given in
Ref.\cite{KoenigPRLEPR03,Vladimirova08} and Ref.\cite{BaratePRB10}. In
particular, Ref.\cite{BaratePRB10} identifies the anticrossing Mn mode as
$\hat{M}_{+,\mathbf{q}}^{(1)}$ consistent with the homothetic approximation of
Eq. (\ref{homothetic}) used here. The anticrossing gap was found to be
$\Delta\omega_{q=0}=\sqrt{4K\Delta_{2}-\left(  \frac{\hbar}{T_{2e}}\right)
^{2}}$ where $\frac{\hbar}{T_{2e}}$ is the damping rate of the homogeneous
uncoupled electron mode, a quantity that we have neglected here in dropping
electron-Mn correlation terms contained in Eqs.(\ref{HmixedL})-(\ref{Hcorrel}%
). In Ref.\cite{BaratePRB10}, $\frac{\hbar}{T_{2e}}$ was estimated from
measurements of the electronic spin wave damping at $q=0$. A rigorous
simultaneous determination of $\Delta_{2}$ and $T_{2e}$, lead to the
extraction of $K$ from the anticrossing gap and furthermore to the
spin-polarization degree $\zeta$ of the 2DEG. Data showed that the
so-extracted $\zeta$ was slightly exceeding the prediction\cite{SpinReponse}
made for $\zeta$ in contradiction with other determinations of $\zeta$
performed in the same type of samples\cite{AkulehPRB07}, which showed that the
model used to predict $\zeta$was reliable. A more accurate description of the
anticrossing gap taking into account the infinite set of coupled $n$-profile
mode $\hat{M}_{+,\mathbf{q}}^{(n)}$ equation of motion might overcome this
discrepancy. One should also mention, that the infinite serie of equations
must be cut in order to conserve the initial number of degrees of freedom
(number of available spins in the system). But finding the right number of Eq.
(\ref{Mndynamixed})-like to be taken depends on how the total number of
degrees of freedom separates into a number of (quasi-) individual
modes\cite{Vladimirova08} and a number of collective electron-Mn modes.
Determining this separation is also an important and interesting issue.

\subsection{Spin waves}%

%TCIMACRO{\FRAME{fhFU}{6.0502in}{4.2566in}{0pt}{\Qcb{(Color online).
%Illustration of the changes introduced by the Coulomb interaction on the mixed
%modes dispersions for the same sample parameters as in Fig.\ref{FigAntiCross}.
%(a) Solutions without Coulomb given by Eq.(\ref{NonIntProp}) for $B=5.8$ T. At
%$q=0$ SF-SPE are degenerate to $Z_{e}$. The $sd$ dynamical shift introduces
%two propagating waves : above (resp. below) the SF-SPE domain, the OPW (resp.
%IPW) propagates with a positive (resp. negative) dispersion. The dashed line
%is the uncoupled Mn mode (degenerate to $\hbar\omega_{Mn}$). (b) and (c) The
%Coulomb interaction between electrons is included as in Eq.(\ref{Mnpoles}) of
%this work. Dashed lines are the uncoupled electron and Mn modes. At $q=0$ the
%SF-SPE energy is shifted to $Z^{\ast}$. An anticrossing gap opens at the
%wavevector $q_{\text{R}}\left(  B\right)  $ if $B\leq B_{\text{R}}.$ In ususal
%conditions (see Ref.\cite{xCxsd}) both the OPW and IPW propagate below the
%SF-SPE continuum with negative dispersions. Dipersions were calculated after
%setting to zero the kinetic damping rate.}}{\Qlb{FigCoulomb}}{Figure}%
%maintain-aspect-ratio TRUE;  display "USEDEF";  valid_file "T";
%width 6.0502in;  height 4.2566in;  depth 0pt;  original-width 4.5204in;
%original-height 3.1739in;  cropleft "0";  croptop "1";  cropright "1";
%tempfilename 'FigDispCoulombyesno.wmf';tempfile-properties "XNPR";}}}%
%BeginExpansion
\begin{figure}
[h]
\begin{center}
\includegraphics[
natheight=3.173900in,
natwidth=4.520400in,
height=4.2566in,
width=6.0502in
]%
{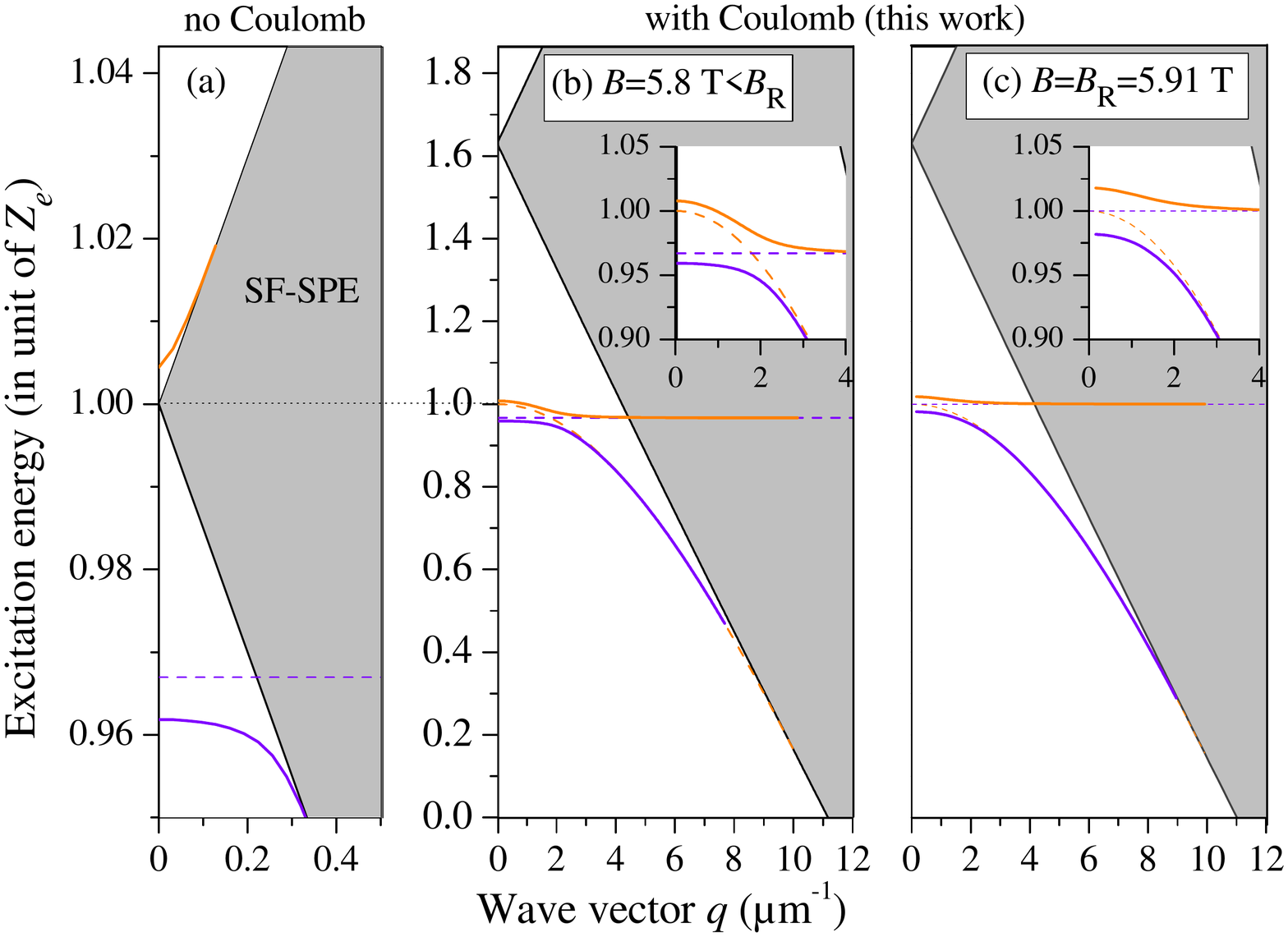}%
\caption{(Color online). Illustration of the changes introduced by the Coulomb
interaction on the mixed modes dispersions for the same sample parameters as
in Fig.\ref{FigAntiCross}. (a) Solutions without Coulomb given by
Eq.(\ref{NonIntProp}) for $B=5.8$ T. At $q=0$ SF-SPE are degenerate to $Z_{e}%
$. The $sd$ dynamical shift introduces two propagating waves : above (resp.
below) the SF-SPE domain, the OPW (resp. IPW) propagates with a positive
(resp. negative) dispersion. The dashed line is the uncoupled Mn mode
(degenerate to $\hbar\omega_{Mn}$). (b) and (c) The Coulomb interaction
between electrons is included as in Eq.(\ref{Mnpoles}) of this work. Dashed
lines are the uncoupled electron and Mn modes. At $q=0$ the SF-SPE energy is
shifted to $Z^{\ast}$. An anticrossing gap opens at the wavevector
$q_{\text{R}}\left(  B\right)  $ if $B\leq B_{\text{R}}.$ In ususal conditions
(see Ref.\cite{xCxsd}) both the OPW and IPW propagate below the SF-SPE
continuum with negative dispersions. Dipersions were calculated after setting
to zero the kinetic damping rate.}%
\label{FigCoulomb}%
\end{center}
\end{figure}
%EndExpansion

For $q>0,$ Eqs(\ref{Mnpoles}) and (\ref{NonIntProp}) give very different
qualitative results as illustrated on Fig.~\ref{FigCoulomb}. Without Coulomb
interaction between electrons, uncoupled modes of the electrons are the SF-SPE
which are degenerate to $Z_{e}$ at $q=0.$ The $sd$ dynamical coupling
introduces two additional collective mode : the OPW propagating above the
SF-SPE domain with a positive dispersion and the IPW propagating below with a
negative dispersion. Introducing the Coulomb interaction between electrons
shifts the SF-SPE to higher energies ($Z^{\ast})$ and givs rise to the
collective wave SFW propagating below the SF-SPE continuum. The SFW is further
coupled to Mn modes through the $sd$ interaction. An evaluation of the
coupling between SF-SPE and Mn modes was given in Ref.\cite{BaratePRB10} and
found to be negligible. Thus the Coulomb interaction introduces a shift
between the SF-SPE and the SFW energies, and the later is further shifted by
the $sd$ dynamical coupling. In realistic conditions, it was shown in
Ref.\cite{xCxsd} that the Coulomb shifts dominates over the $sd$ dynamical
shift. Hence, when Coulomb interaction is taken into account, Eq.
(\ref{Mnpoles}), except under unrealistic conditions, gives rise to two spin
wave modes propagating below the SF-SPE continuum, the IPW and OPW. An
anticrossing gap opens at a specific wavevector $q_{\text{R}}\left(  B\right)
$ given by :%

\begin{equation}
q_{\text{R}}\left(  B\right)  =\sqrt{\left\vert \zeta\right\vert \left(
Z^{\ast}/Z_{e}-1\right)  \frac{2m^{\ast}}{\hbar}\left(  \omega_{e}-\omega
_{Mn}\right)  } \label{qR}%
\end{equation}

Note that $q_{\text{R}}\left(  B_{\text{R}}\right)  =0$ and that
$\tilde{\omega}_{q_{\text{R}}}=\omega_{Mn}+iq_{\text{R}}^{2}G_{xc}\sigma_{+}.$
If $q_{\text{R}}\left(  B\right)  >0$, compared to the homogenous gap, the
anticrossing gap at $q_{\text{R}}$ is\ dramatically reduced by the kinetic
damping of the electron wave and is given by :%

\begin{equation}
\omega_{q_{\text{R}}}^{+}-\omega_{q_{\text{R}}}^{-}=\sqrt{4K\Delta_{2}%
/\hbar^{2}-\eta_{q_{\text{R}}}^{2}} \label{GapR}%
\end{equation}

where $\eta_{q_{R}}=q_{\text{R}}^{2}G_{xc}\sigma_{+}.$ We note that the
kinetic damping is the only one considered here. Other sources of damping, as
e.g., the ones dropped in Eq. (\ref{Hcorrel}), will of course further reduce
the amplitude of the gap.%

%TCIMACRO{\FRAME{fhFU}{6.0502in}{4.2557in}{0pt}{\Qcb{(Color online). Left axis
%: variation of the wavevector $q_{\text{R}}\left(  B\right)  $ and the
%wavevector $q_{m}=k_{F\downarrow}-k_{F\uparrow}.$ Right axis : variation of
%the anticrossing gap and the kinetic damping rate $\eta_{q_{\text{R}}%
%}=\operatorname{Im}\tilde{\omega}_{q_{R}}$. The anticrossing gap is killed by
%the damping rate when the later is of the same order of magnitude as
%$\sqrt{4K\Delta_{2}}$. A break in the \ horizontal axis scale has been
%introduced to zoom the region close to $B\precsim B_{\text{R}}.$ To calculate
%the damping rate, we have used a typical SF-SPE scattering time $\tau=2$ps
%instead of $\tau_{e-e}$ ($\sim150$ps) to match the experimental conditions of
%Ref.\cite{Gomez}.}}{\Qlb{figGapR}}{Figure}%
%maintain-aspect-ratio TRUE;  display "USEDEF";  valid_file "T";
%width 6.0502in;  height 4.2557in;  depth 0pt;  original-width 4.5204in;
%original-height 3.1739in;  cropleft "0";  croptop "1";  cropright "1";
%cropbottom "0";  tempfilename 'FigGapR.wmf';tempfile-properties "XNPR";}}}%
%BeginExpansion
\begin{figure}
[h]
\begin{center}
\includegraphics[
natheight=3.173900in,
natwidth=4.520400in,
height=4.2557in,
width=6.0502in
]%
{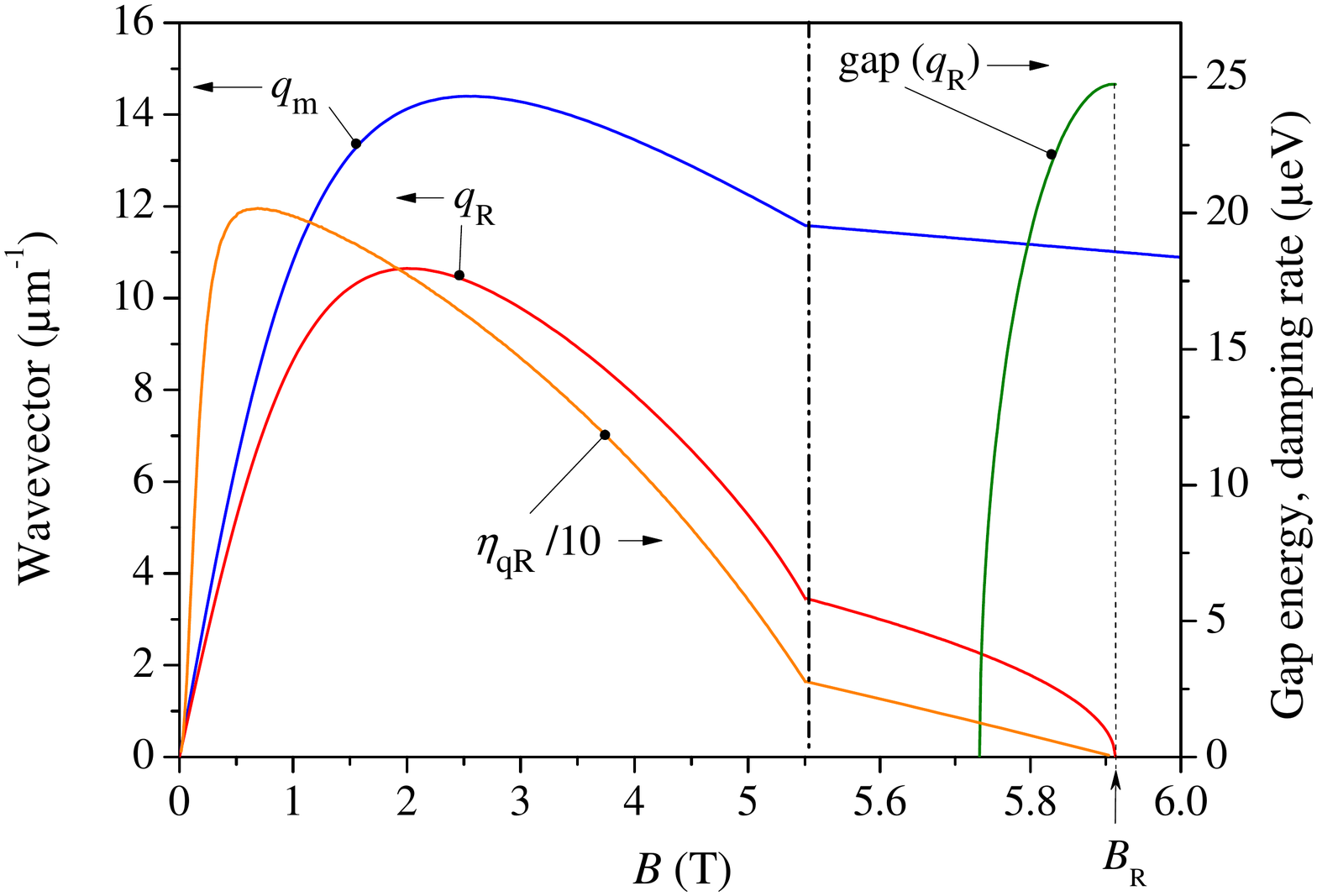}%
\caption{(Color online). Left axis : variation of the wavevector $q_{\text{R}%
}\left(  B\right)  $ and the wavevector $q_{m}=k_{F\downarrow}-k_{F\uparrow}.$
Right axis : variation of the anticrossing gap and the kinetic damping rate
$\eta_{q_{\text{R}}}=\operatorname{Im}\tilde{\omega}_{q_{R}}$. The
anticrossing gap is killed by the damping rate when the later is of the same
order of magnitude as $\sqrt{4K\Delta_{2}}$. A break in the \ horizontal axis
scale has been introduced to zoom the region close to $B\precsim B_{\text{R}%
}.$ To calculate the damping rate, we have used a typical SF-SPE scattering
time $\tau=2$ps instead of $\tau_{e-e}$ ($\sim150$ps) to match the
experimental conditions of Ref.\cite{Gomez}.}%
\label{figGapR}%
\end{center}
\end{figure}
%EndExpansion

Fig.~\ref{figGapR} illustrates the variation of $q_{\text{R}}\left(  B\right)
$ with the magnetic field. It is always smaller than $q_{m}=k_{F\downarrow
}-k_{F\uparrow}$, the wavevector delimiting the window where the SFW
propagates\cite{SpinReponse}. Overlaid in Fig.~\ref{figGapR}, are the
anticrossing gap at $q_{\text{R}}$ and the corresponding damping rate
$\eta_{q_{\text{R}}}.$ In the absence of the kinetic damping, the gap would be
given by $\sqrt{4K\Delta_{2}}.$ One sees the dramatic effect of this intrinsic
kinetic damping, which kills the gap outside a very narrow range of magnetic
fields. As coupling between spin waves of the electron and the Mn spin systems
is responsible for the appearance of the carrier induced
ferromagnetism\cite{KoenigPRLEPR03}, we might conclude that the above
disappearance of the gap diminishes the possibilities for ferromagnetic
transitions with complex order (out of $q=0$).%

%TCIMACRO{\FRAME{fhFU}{6.0494in}{4.2748in}{0pt}{\Qcb{(Color online).
%Illustration of the progressive disappearance of the anticrossing gap when $B$
%goes away from $B_{R}.$ Continuous lines (open symbols) are dispersions
%calculated in absence (presence) of the kinetic damping. Dashed lines are the
%uncoupled modes. Upper insets : zoom on the anticrossing gap region. Lower
%insets : variation with $q$ of the damping rate of both the OPW (dashed line)
%and IPW(straight line).}}{\Qlb{FigGapDisappear}}{Figure}%
%maintain-aspect-ratio TRUE;  display "USEDEF";  valid_file "T";
%width 6.0494in;  height 4.2748in;  depth 0pt;  original-width 4.5204in;
%original-height 3.1739in;  cropleft "0";  croptop "1";  cropright "0.9956";
%tempfilename 'FigGapDisappear.wmf';tempfile-properties "XNPR";}}}%
%BeginExpansion
\begin{figure}
[h]
\begin{center}
\includegraphics[
trim=0.000000in 0.000000in 0.019890in 0.000000in,
natheight=3.173900in,
natwidth=4.520400in,
height=4.2748in,
width=6.0494in
]%
{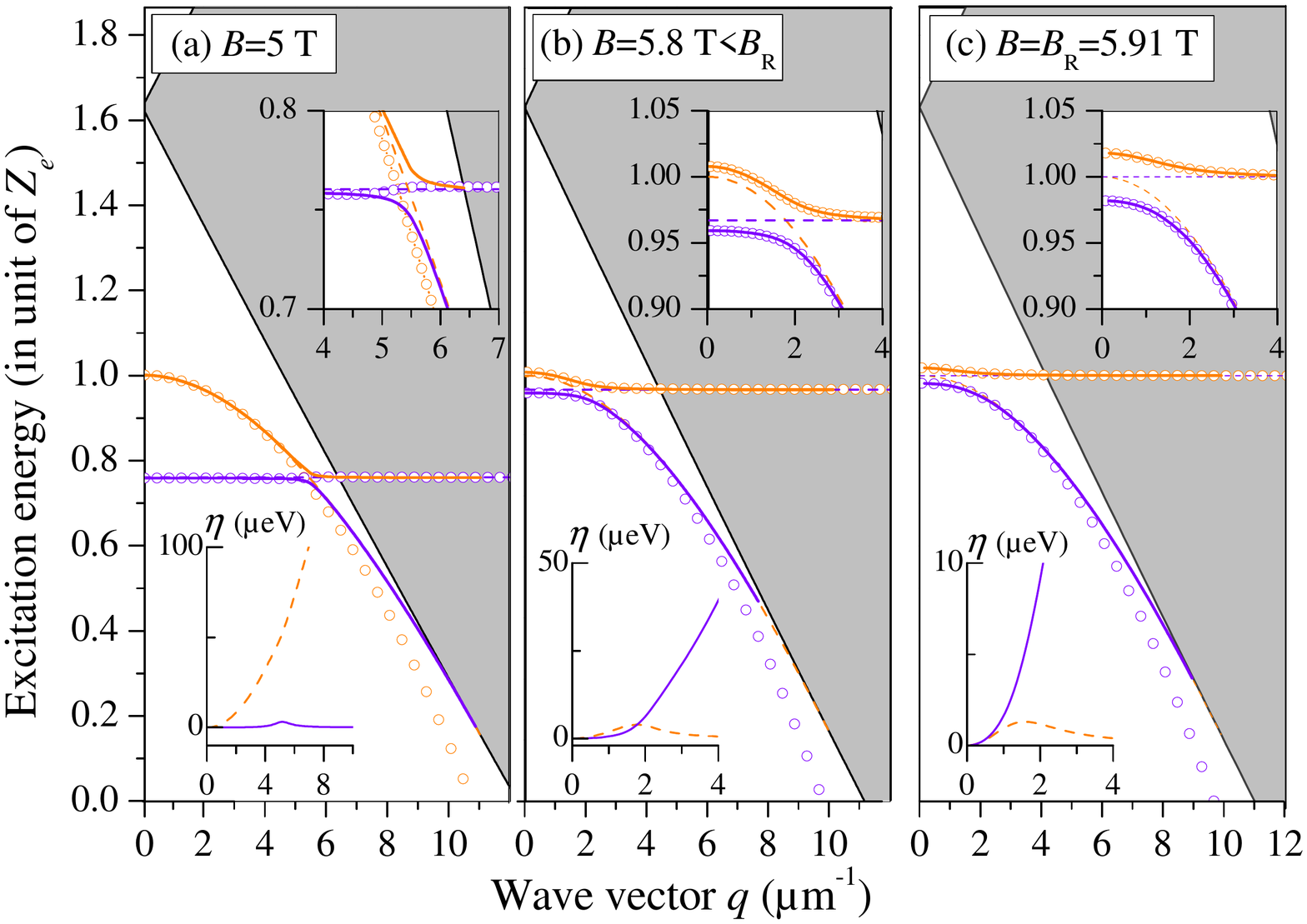}%
\caption{(Color online). Illustration of the progressive disappearance of the
anticrossing gap when $B$ goes away from $B_{R}.$ Continuous lines (open
symbols) are dispersions calculated in absence (presence) of the kinetic
damping. Dashed lines are the uncoupled modes. Upper insets : zoom on the
anticrossing gap region. Lower insets : variation with $q$ of the damping rate
of both the OPW (dashed line) and IPW(straight line).}%
\label{FigGapDisappear}%
\end{center}
\end{figure}
%EndExpansion

The disappearance of the gap is illustrated in Fig.~\ref{FigGapDisappear} by
comparing the dispersions obtained from the zeros of the propagator in Eq.
(\ref{Mnpoles}) in the presence or absence of the kinetic damping. In the
presence of the damping, the solutions have a non-zero imaginary part for
$q>0.$ The corresponding damping rate is plotted in the lower insets of
Fig.~\ref{FigGapDisappear}. It is well known that when the frequencies of two
coupled oscillators anticross each other, their corresponding damping rates
cross themselves. Clearly for $B<5.7$ T, the mixed modes do not anticross at
any $q$ and each branch conserves its former character, Mn-like or
electron-like. On the contrary, for $5.7$ T$<B<B_{\text{R}},$ the modes
anticross at $q_{\text{R}}$, and the OPW transfers the kinetic damping
($q^{2}$ law) of the SFW to the IPW when $q>q_{\text{R}}$. It is worth to note
that this $q^{2}$ law for the IPW damping rate was also found in GaMnAs
compounds in the ferromagnetic state\cite{PerakisPRL08}.

\bigskip\bigskip

\section{Conclusion}

In conclusion, we have introduced equations for the spin dynamics in a
test-bed diluted magnetic system that allow to take into account the interplay
between the Coulomb interaction dynamics and the $sd$ dynamical coupling
between the electrons and the localized spins. We have shown how the Coulomb
interaction introduces strong qualitative changes : the mixed electron-Mn
modes propagate below the SF-SPE continuum and an anticrossing gap is open for
a given range of magnetic field. Because of Coulomb interaction, the intrinsic
kinetic damping due to the electron motion is always present (the SF-SPE
scattering time can not be longer than $\tau_{e-e}$), this damping kills the
anticrossing gap outside a very narrow range of magnetic fields. Our
calculations illustrate also how this kinetic damping is transfered to the
IPW, a phenomenon found in GaMnAs compounds.

\begin{acknowledgments}
The authors would like to thank I. d'Amico, E. Hankiewicz and the GOSPININFO
consortium for fruitfull discussions as well as the grant ANR GOSPININFO for
financial support.
\end{acknowledgments}

\end{document}